\documentclass[prb]{revtex4-1} 

\usepackage{amsmath} % needed for \tfrac, \bmatrix, etc.
\usepackage{amsfonts} % needed for bold Greek, Fraktur, and blackboard boldV\usepackage{graphicx} % needed for figures
\usepackage{epstopdf}
\usepackage{algorithmicx}
\usepackage{algcompatible}
\usepackage{algorithm} %ctan.org\pkg\algorithms

\usepackage{dcolumn}% Align table columns on decimal point
\usepackage{bm}% bold math
\DeclareMathOperator\arctanh{arctanh}
\usepackage{hyperref}% add hypertext capabilities
\usepackage{graphics}
\usepackage{epsfig}
\usepackage{bbold}
\usepackage{verbatim}

\newcommand{\EQ}[1]{Eq.~(\ref{eq:#1})}
\newcommand{\EQS}[2]{Eqs.~(\ref{eq:#1}) and (\ref{eq:#2})}
\newcommand{\FIG}[1]{Fig.~\ref{fig:#1}}

\def\bal#1\eal{\begin{align}#1\end{align}}
\newcommand{\be}{\begin{equation}}
\newcommand{\ee}{\end{equation}}
\newcommand{\ba}[1]{\begin{array}{*{#1}{c}}}
\newcommand{\ea}{\end{array}}

\begin{document}

\title{Lifting -- A nonreversible Markov chain Monte Carlo algorithm}

\author{Marija Vucelja}
\email{mvucelja@virginia.edu}
\affiliation{Center for Studies in Physics and Biology, The Rockefeller University,
1230 York Avenue, New York, NY, 10065, USA}
\affiliation{Department of Physics, University of Virginia, Charlottesville, 22904, VA, USA}

\date{\today}

\begin{abstract}
Markov chain Monte Carlo algorithms are invaluable tools for exploring stationary properties of physical systems, especially in situations where direct sampling is unfeasible. Common implementations of Monte Carlo algorithms employ reversible Markov chains. Reversible chains obey detailed balance and thus ensure that the system will eventually relax to equilibrium. Detailed balance is not necessary for convergence to equilibrium. We review nonreversible Markov chains, which violate detailed balance, and yet still relax to a given target stationary distribution. In particular cases, nonreversible Markov chains are substantially better at sampling than the conventional reversible Markov chains with up to a square root improvement in the convergence time to the steady state. One kind of nonreversible Markov chain is constructed from the reversible ones by enlarging the state space and by modifying and adding extra transition rates to create non-reversible moves. Because of the augmentation of the state space, such chains are often referred to as \emph{lifted} Markov Chains. We illustrate the use of lifted Markov chains for efficient sampling for several examples. The examples include sampling on a ring, sampling on a torus, the Ising model on a complete graph, and the one-dimensional Ising model. We also provide a pseudocode implementation, review related work, and discuss the applicability of such methods.
\end{abstract}

\maketitle 
\section{Monte Carlo -- an invaluable method}
We resort to Monte Carlo algorithms when faster converging numerical approaches are inapplicable. Such is usually the case in statistical physics and in quantum field theory, where we often need to evaluate high-dimensional integrals. For example, a well-known discretization technique, Simpson's rule\cite{integralhouse} estimates a $d$-dimensional integral by partitioning it into $n$ segments with an error proportional to $n^{-4/d}$. In comparison, for the majority of Monte Carlo methods, the error scales as $n^{-1/2}$ and more importantly the error is independent of the dimension. Already for $d>8$ the Monte Carlo method outperforms Simpson's rule. However, Monte Carlo is still rather slow, and thus one should think of Monte Carlo as a useful last resort -- as Sokal in his lecture notes cautions:\cite{97Sokal} ``Monte Carlo is an extremely bad method; it should be used only when all alternative methods are worse.''

In statistical physics, we typically work with systems that can occupy exponentially many states (for example, $2^N$ states of $N$ Ising spins and $4^N$ genomic sequences of length $N$). Because direct sampling across the enormous phase space is unfeasible, dynamical Monte Carlo algorithms are often the only choice. In a dynamical Monte Carlo method, we define a stochastic process on the configuration space of the physical system such that as time goes to infinity, the process relaxes to equilibrium. Due to its simplicity and being ``memoryless'' (the time evolution depends only on the present) a common choice for the stochastic process is a \emph{Markov chain}. We will focus on dynamical Monte Carlo methods utilizing Markov chains, known as \emph{Markov chain Monte Carlo} methods. At sufficiently large times the system will be close enough to equilibrium so that we can compute equilibrium average physical quantities of interest (such as the magnetization in a spin system, spin-spin correlation function, partition function, susceptibility, and conductivity). This stochastic time evolution of the Markov chain from an arbitrary initial distribution to the vicinity of the target (equilibrium) distribution is a fictitious auxiliary time evolution, and much work is being done to speed up its convergence.

The focus of our article is to show how to modify the stochastic time evolution to accelerate the convergence to equilibrium. Our intuition stems from hydrodynamics and chaotic mixing. We will make use of fluid dynamics analogies when we introduce our nonreversible Markov chain Monte Carlo method.\cite{turitsyn2011irreversible} One way to introduce a nonreversible algorithm is to modify the corresponding reversible Markov chain Monte Carlo by allowing for nonreversible moves over a phase space that has been enlarged from the original one. In mathematics and computer science literature, these kinds of nonreversible Markov chain Monte Carlo methods are called \emph{lifted Markov chain Monte Carlo} methods.\cite{CLP00,HS} 

In Sec.~\ref{sec:pre} we explain the general idea behind dynamical Monte Carlo methods, the relevant mathematical prerequisites, and our notation. In Sec.~\ref{sec:mh} we describe a famous variant of the Markov chain Monte Carlo algorithm family -- the Metropolis-Hastings algorithm, followed in Sec.~\ref{sec:converge} by measures of relaxation to equilibrium. In Sec.~\ref{sec:lift} we introduce the lifted Markov chain Monte Carlo method. We illustrate it with pertinent examples in Sec.~\ref{sec:appl_lifting} and conclude with a discussion in Sec.~\ref{sec:discussion}. Suggested problems are given in Sec.~\ref{sec:sugg_problems}. 

\section{Dynamical Monte Carlo, mathematical prerequisites and notation}
\label{sec:pre}
Consider a system prepared in an initial state $x$, chosen from a finite set of possible states $\Omega$ and suppose we know the equilibrium probability distribution $\rho_{\rm equil} (x)$. The states can represent spin configurations, particle locations and velocities, polymer conformations, genetic sequences, for example. Our goal is to determine various macroscopic statistical properties of the system, such as thermodynamic averages of observables such as the mean energy, magnetization, and mean time to a common ancestor. For a very large state space $\Omega$ it is impossible to visit all of the states in real time. Likewise, it is unfeasible to evaluate the thermodynamic average of an observable, denoted here by $f$, from the definition
\bal
\label{eq:ensemble_average}
\langle f \rangle_{\rho_{\rm equil} } \equiv \sum _{x\in \Omega} \rho_{\rm equil} (x) f(x). 
\eal
Instead, we can use a dynamical Monte Carlo algorithm and define a stochastic process on $\Omega$ with a stationary distribution $\rho_{\rm equil}$. This stochastic process is defined so that it visits more often states that are more probable, than those that are less probable in equilibrium. At large enough times the histogram of the visited states gives a numerical estimate of the limiting distribution; the ``visiting bias'' of the stochastic process is adjusted so that the limiting distribution is $\rho_{\rm equil}$. In other words, by using such a stochastic process we efficiently sample the phase space, and the time-evolution the stochastic process brings the probability distribution ever closer to $\rho_{\rm equil}$ on average.

Typically a Markov chain is chosen for the dynamical Monte Carlo stochastic process. A \emph{Markov chain} is a sequence of states for which the probability of the next state is fully determined by the present state and is independent of the past path. This ``memoryless'' property is known as the \emph{Markovian property}. That is, in a Markov chain the conditional probability of going from state $x$ to state $y$ at the next time step, $P(x,y)$, is independent of the path that took the system to state $x$. This ``independence from the past'' is the reason that a matrix of size $|\Omega|\times|\Omega|$ is sufficient to specify the evolution of the system. 

The $x$th row of the transition matrix $P$ is itself a distribution. The matrix $P$ has non-negative elements, and as a corollary of the conservation of probability 
\be
\label{eq:P_stochastic}
1 = {\rm Prob}(x \to x\text{ or any other state in }\Omega) = \sum_{y\in \Omega} P(x,y),
\ee 
holds for all $x\in \Omega$. We call such a matrix \emph{stochastic}. An initial probability distribution $\rho (t = 0)$ evolves to the probability distribution $\rho(t)$ according to
\be
\rho(t;x) = \sum_{y\in\Omega} \rho(t;y)P(y, x) = \ldots = \sum_{y\in\Omega}\!\rho(0;y)P^t(y, x),
\ee
or using vector notation 
\bal
\rho(t) = \rho(t-1)P =\rho(t-2)P^2= \ldots =\rho(0)P^t,
\eal
where the time $t \geq 0$ is assumed to be discrete and measured in the number of steps (number of transitions attempted by the Markov chain), and $P^t$ is the matrix $P$ raised to the power $t$. Note that here we are multiplying a matrix by a vector on the left. Continuous time Markov chains can also be defined, but are beyond the scope of our review.

A probability distribution $\pi$ is \emph{stationary} if it does not change with time, that is,
\be
\label{eq:stationary_dist}
\pi = \pi P. 
\ee
A transition matrix $P$ is \emph{irreducible} if a path can be found between any two states on $\Omega$; that is, for any two states $x$ and $y$ in $\Omega$ there exists an integer $t$ such that $P^t(x,y)>0$. The irreducibility of $P$ means that it is possible to go from any state to any other state using only transitions of non-zero probability; physicists usually call this property \emph{ergodicity}. A transition matrix $P$ is \emph{aperiodic} if the greatest common divisor of a set of times $\mathcal{T}(x) = \{t \geq 1 : P^t (x,x) >0, \forall x \in \Omega \}$ is 1.

\begin{figure}
\includegraphics[width=3.42in]{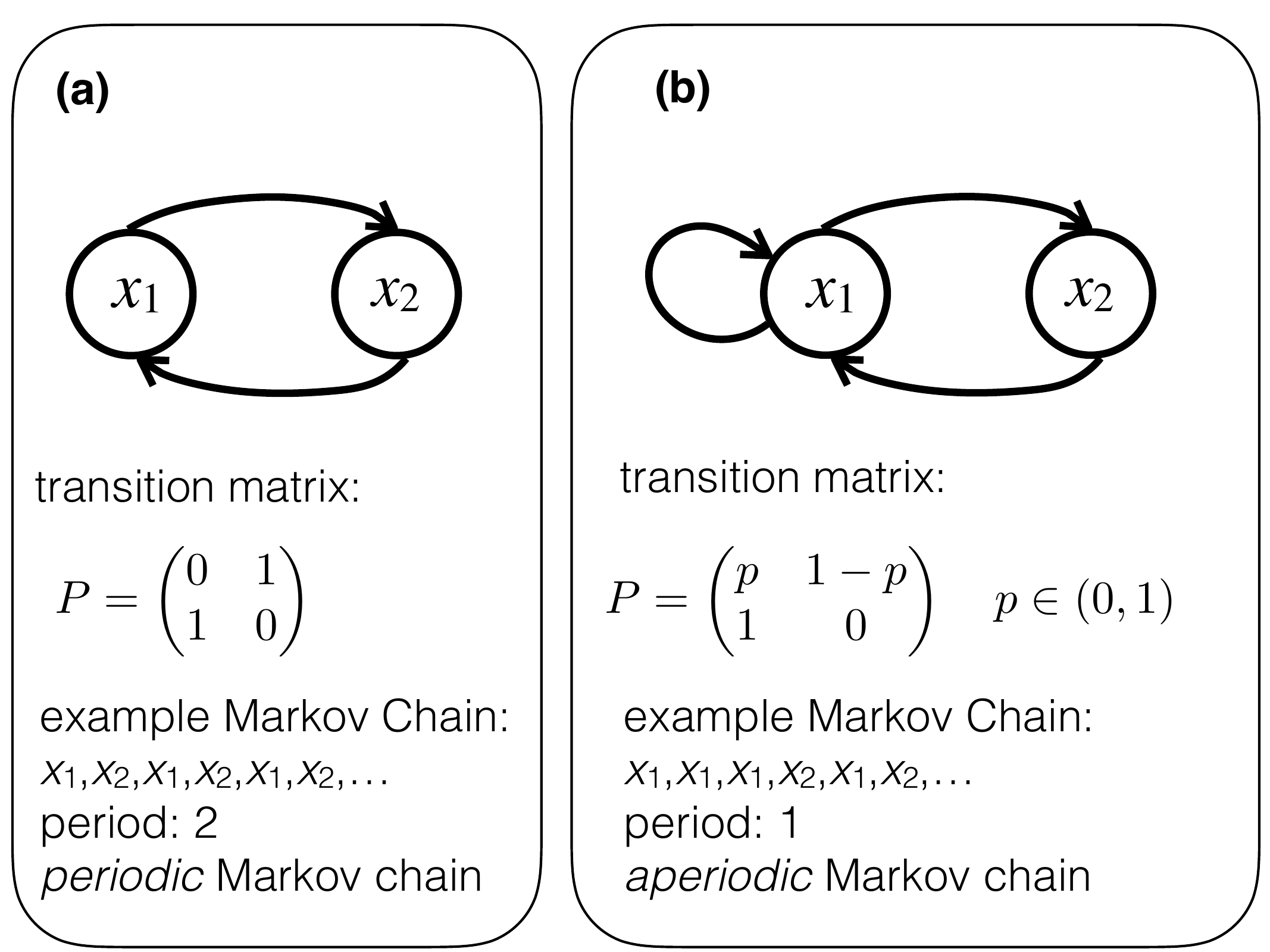}
\caption{\label{fig:aperiodicity.pdf} (a) An example of a graph that has a periodic Markov chain. (b) An example of a graph with an aperiodic Markov chain.}
\end{figure}

To understand aperiodicity let us look at two graphs, depicted in \FIG{aperiodicity.pdf}, each consisting of only two states, labelled $x_1$ and $x_2$. The graph in \FIG{aperiodicity.pdf}(a) has a periodic transition matrix, because the possible Markov chains are alternating sequences determined by the initial condition: $\{x_1,x_2,x_1,\ldots\}$ or $\{x_2,x_1,x_2,\ldots\}$. The period of returning to the same state is $1$. The graph in \FIG{aperiodicity.pdf}(b) has an aperiodic $P$; there is no clear period of returning to $x_1$ or $x_2$. An example of a Markov chain, starting from $x_1$ is: $\{x_1,x_1,x_2,x_1,x_2,x_1,x_1,x_1,\ldots\}$.

From \EQS{P_stochastic}{stationary_dist} we see that a Markov chain $P$ that has $\pi$ as a stationary distribution satisfies a global balance condition 
\bal
\label{eq:GBC}
\sum_{y \in \Omega} \left[\pi(x) P(x,y) - \pi(y) P(y,x)\right]=0\,, \quad \forall x \in \Omega.
\eal

In the following we will focus on irreducible, aperiodic and finite (defined over finite state spaces) Markov chains.\cite{2009Levinbook} For such Markov chains, the stationary distribution, if it exists, is unique (see, for example, Refs.~\onlinecite{97Sokal, 2009Levinbook}).

The global balance condition, \EQ{GBC}, signifies that the total influx to a state is equal to the total efflux from this state. That is, to employ a hydrodynamics analogy, global balance amounts to the incompressibility of phase space. A special case of \EQ{GBC} is the pairwise cancelation of the terms in the sum; it is called \emph{the detailed balance condition}:
\bal
\label{eq:DBC}
\pi(x) P(x,y) = \pi(y) P(y,x)\,, \quad \forall x,y\in\Omega. 
\eal
In contrast to global balance, detailed balance is a local, microscopic reversibility property. A hydrodynamic analogy is an irrotational flow. Detailed balance is a special case of global balance, just as all irrotational flows are incompressible. Markov chains obeying detailed balance are called reversible Markov chains. It is usually much easier to implement detailed balance because it is a local condition.

\section{Metropolis-Hastings algorithm}
\label{sec:mh}
One of the most famous Markov chain Monte Carlo algorithms is due to Metropolis et al.\cite{MRRTT53} It is called the \emph{Metropolis-Hastings algorithm}, because it was later generalized by Hastings.\cite{H70} We will use the Metropolis-Hastings algorithm in all of our examples that follow.

We start by explaining how the Metropolis-Hastings algorithm works. We are given the target equilibrium distribution $\pi$, but because the space of states $\Omega$ is large, we cannot find averages directly from the definition, \EQ{ensemble_average}. In the Metropolis-Hastings algorithm, a Markov chain is used as a stochastic process. Next we need to ensure that the Markov chain dynamics relaxes the system to equilibrium, that is, that the limiting distribution of the Markov chain is indeed $\pi$. We start with an arbitrary initial distribution $\rho(t=0)$ and an arbitrary transition matrix $Q$, which specifies a Markov chain with a stationary distribution, which usually is different from $\pi$. Our objective is to find a new transition matrix $P$, such that the new Markov chain $P$ has as its stationary distribution $\pi$. This goal can be obtained by introducing \emph{acceptance probabilities} $a(x,y)$ in such a way that the resulting transition matrix, $P$, with off-diagonal elements ($x\neq y$)
\bal
P(x,y) = a(x,y)Q(x,y),
\eal
obeys detailed balance. The diagonal elements of $P$ are set by conservation of total probability: $P(x,x) = 1 - \sum_{y\in \Omega \backslash x} P(x,y)$. As we noted in Sec.~\ref{sec:pre} an aperiodic, irreducible, Markov chain $P$ that obeys detailed balance, \EQ{DBC}, at large times has $\pi$ as its stationary distribution. For \EQ{DBC} to hold, the acceptance probability should satisfy 
\bal
\label{eq:acceptance_criterion}
\frac{a(x,y)}{a(y,x)} = \frac{\pi(y)Q(y,x)}{\pi(x)Q(x,y)}. 
\eal
Solutions of \EQ{acceptance_criterion} include the \emph{heat bath acceptance probability} 
\bal
\label{eq:heat_bath_acceptance_probability}
&a(x,y) = \frac{\pi(y)Q(y,x)}{\pi(y)Q(y,x)+\pi(x)Q(x,y)},
\eal
and the \emph{Metropolis-Hastings acceptance probability}
\bal
\label{eq:Metropolis-Hastings_acceptance_probability}
&a(x,y) = \min \left[ 1, \frac{\pi(y)Q(y,x)}{\pi(x)Q(x,y)}\right]. 
\eal
For the special case of a symmetric transition matrix $Q$, $Q(x,y) = Q(y,x)$, the Metropolis-Hastings acceptance probability simplifies to 
\bal
&a(x,y) = \min \left[ 1, \frac{\pi(y)}{\pi(x)}\right].
\eal
Notice that if $y$ is more probable than $x$, that is $\pi(y) > \pi(x)$, the proposed move $x \to y$ is always accepted ($a(x,y) = 1$). This is the ``visiting bias'' that we mentioned earlier. Both acceptance probabilities, \EQS{heat_bath_acceptance_probability}{Metropolis-Hastings_acceptance_probability}, are widely used.

Now that we have defined a correct Markov chain Monte Carlo process, in that it converges to the given target distribution $\pi$, we need to know how fast it converges, which is the topic of Sec.~\ref{sec:converge}. 

\section{\label{sec:converge}Convergence measures}
We are typically interested in the equilibrium average values and equilibrium correlations between observables. Following Ref.~\onlinecite{97Sokal}, we will describe various ways to measure the relaxation to equilibrium. Suppose $f$ is an observable that is a function of the possible system states. For example, in a magnetic system, the magnetization is a function of the spin configuration, as is the energy of the system. We define a Markov process with a transition matrix $P$ and start the system with the initial distribution $\rho(t=0)$. The mean $\mu_f(t)$ and variance $\sigma^2_f$ of the observable $f$ are time dependent and equal to 
\bal
\mu_f(t) &\equiv \langle f(t)\rangle \equiv \sum_{x\in \Omega} \rho(t;x)f(x),
\\
\sigma^2_f(t) & \equiv \langle f^2(t)\rangle - \langle f(t)\rangle^2 = \sum_{x\in\Omega}\rho(t;x)[f(x) - \mu_f(t)]^2.
\eal
For the types of Markov processes that we will consider, $\rho(t;x)$ converges to the equilibrium distribution $\pi$ and the average properties become stationary (time-independent) as time goes to infinity:
\bal
\mu_{f,\pi} & \equiv\langle f\rangle_\pi \equiv \sum_{x\in\Omega}\pi(x)f(x), \\
\sigma^2_{f,\pi} &\equiv \langle f^2 \rangle_\pi - \langle f \rangle^2_{\pi}\equiv \sum_{x\in\Omega}\pi(x)[f(x) - \mu_f]^2.
\eal
We have omitted the dependence on time to stress that these averages are time-independent and added a subscript $\pi$ to emphasize that these are equilibrium averages.

A good measure of how close the system is to equilibrium is the autocorrelation function, which describes the correlations between the stochastic observable at different times. The autocorrelation function $R_{f\!f}(t_1, t_2)$ for the observable $f$ is defined as 
\bal
R_{f\!f}(t_1,t_2) \equiv
\frac{\langle [f(t_1) - \mu_f(t_1)][f(t_2) - \mu_f(t_2)]\rangle}{\sigma_f(t_1)\sigma_f(t_2)}\,. 
\label{rff}
\eal 
For a second-order stationary stochastic process (a process where the first and the second moment do not vary with respect to time) the autocorrelation function depends only on the time difference $t = t_2-t_1$. In this case Eq.~(\ref{rff}) simplifies to 
\bal
\label{eq:autocorr_norm}
R_{f\!f}(t) = 
\frac{\langle f(0)f(t)\rangle_{\pi} - \langle f(0)\rangle^2_\pi}{\langle f(0)^2\rangle_\pi - \langle f(0)\rangle ^2_\pi} =\frac{\langle f(0)f(t)\rangle_\pi - \mu^2_{f,\pi}}{\sigma_{f,\pi}^2},
\eal
where 
$\langle f(0)f(t) \rangle_\pi = \sum_{x,y\in\Omega}f(x)\pi(x)P^t(x,y)f(y)$.

As a measure of convergence to equilibrium it is customary to define the exponential autocorrelation time, the integrated autocorrelation time, and the inverse spectral gap. The exponential autocorrelation time of the observable $f$ is 
\be
\tau_{\exp,f} \equiv \limsup_{t \to \infty} \frac{t}{-\ln|R_{f\!f} (t)|}.
\ee
The time $\tau_{\exp,f}$ is the least upper bound of $t/(-\ln|R_{f\!f}(t)|)$ as $t \to \infty$. 
We define the exponential autocorrelation time $\tau_{\exp}$ to be the relaxation of the slowest observable of the system
\bal
\tau_{\exp} = \sup_f \tau_{\exp,f}. 
\eal
In brief, $\tau_{\exp}$ places an upper bound on the number of iterations that should be discarded at the beginning of a run, before the system is considered to be in equilibrium for all practical purposes. 

The inverse spectral gap and the inverse absolute spectral gap of $P$ are also frequently used as measures of convergence. The transition probability matrix has the following spectral properties: the eigenvalues of the matrix $P$ lie in a unit disk where $\lambda_1 = 1$ is the largest eigenvalue and is non-degenerate (for the chains that we consider: finite, irreducible and aperiodic). The corresponding eigenvector $\phi_1$ is a constant function with $\phi_1(x) = 1$, for all $x\in\Omega$, which follows from stochasticity, \EQ{P_stochastic}. A spectral decomposition of $P$ over the inner product $\langle \phi_i, \phi_j\rangle _\pi = \sum _{x\in \Omega}\phi _i(x)\phi _j(x)\pi (x)$ is
\bal
P(x,y) = \sum ^{|\Omega|} _{j = 1}\phi_j(x)\phi_j(y)\pi(y)\lambda_j.
\eal
If $P$ obeys detailed balance, it has real eigenvalues. To see this we define another matrix $A(x,y) = \pi(x)^{1/2}\pi(y)^{-1/2}P(x,y)$, which has the same eigenvalues as $P$, because 
\bal
A(x,y) = \sqrt{\pi(x)\pi(y)} \left(1+ \sum ^{|\Omega|} _{j = 2} \phi_j(x) \phi_j (y) \lambda_j\right) = \sum ^{|\Omega|} _{j = 1}\left(\frac{\phi_j(x)}{\sqrt{\pi(x)}}\right)\left(\frac{\phi_j(y)}{\sqrt{\pi(y)}}\right) \lambda_j,
\eal
is the spectral decomposition of $A$ over the inner product $\langle \varphi_i, \varphi_j\rangle = \sum _{x\in \Omega}\varphi_i(x)\varphi_j(x)$, where $\varphi_i(x) \equiv\phi_i(x)/\sqrt{\pi(x)}$. We note that the eigenvectors of $A$ and $P$ are different, but the eigenvalues are the same. Finally for the case that $P$ obeys detailed balance we observe that $A$ is symmetric ($A^T =A$):
\bal
A(x,y) = \sqrt{\frac{\pi(x)}{\pi(y)}} P(x,y) = \sqrt{\frac{\pi(x)}{\pi(y)}}\frac{\pi(y)}{\pi(x)}P(y,x) = \sqrt{\frac{\pi(y)}{\pi(x)}}P(y,x) = A(y,x).
\eal
We recall that symmetric matrices have real eigenvalues. Therefore if $P$ obeys detailed balance, it has a real spectrum. 

At a finite time $t$ we have 
\bal
\frac{P^t(x,y)}{\pi(y)} = 1 + \phi_2(x)\phi_2(y)\lambda^t_2 +\mathcal{O}(\lambda^t_3), 
\eal 
where $\phi_2$ is the eigenvector of the second largest eigenvalue. This expression can be written as
\bal
P^t(x,y) &\approx \pi(y)\left[1 + \phi_2(x)\phi_2(y)(1-\Delta)^t\right]\approx \pi(y)\left[1 + \phi_2(x)\phi_2(y)e^{-\Delta t}\right].
\eal
The spectral gap is defined as the difference between the two largest eigenvalues:
\bal
\Delta \equiv \lambda_1-\lambda_2 = 1 - \lambda_2.
\eal 
We see that in the case of real eigenvalues $ \lambda_1$ and $ \lambda_2$, $e^{-\Delta t}$ is a measure of how fast the reversible Markov chain converges to $\pi$.
In contrast, if $P$ obeys global balance, the eigenvalues are in general complex and the system relaxes to equilibrium with damped oscillations. With complex eigenvalues it makes more sense to use the absolute spectral gap $\Delta^* \equiv 1 - |\lambda_2|$ or the real part of the spectral gap, ${\rm Re}(\Delta)$, as measures of convergence. From the definition of $\tau_{\exp}$ it follows that
\bal 
\tau_{\exp} = \frac{1}{\Delta^*}.
\eal In this case a system relaxes to equilibrium with damped oscillations. 

Another useful measure of convergence is the integrated autocorrelation time $\tau_{\rm int}$:
\bal
\label{eq:tau_int}
\tau_{\rm int} \equiv \sup_{f} \tau_{{\rm int},f} = \sup_f \left( \frac{1}{2}+\sum^\infty_{t=1}R_{f\!f}(t)\right).
\eal 
Note, that if $R_{f\!f}(t) \sim e^{-t/\tau}$ and $\tau \gg 1$, we have 
\bal
\tau_{{\rm int},f}\approx \tau_{{\rm \exp}, f},
\eal
which can be checked by direct substitution. The integrated autocorrelation time controls the statistical error in Monte Carlo measurements of equilibrium averages, such as $\mu_{f,\pi}$. 

For further details on Markov chain Monte Carlo methods and convergence see Ref.~\onlinecite{97Sokal}. Another excellent source on Monte Carlo algorithms and their applications to statistical physics is Ref.~\onlinecite{Kra06}.

\section{Lifting}
\label{sec:lift}
Markov chain Monte Carlo methods that obey detailed balance use equilibrium dynamics to sample phase space. For example, consider a phase space lattice with a uniform steady state distribution $\pi$. Each point on the lattice represents a state of the system, and a uniform distribution means that each lattice point is equally likely to be occupied. Metropolis-Hastings moves are unbiased hops to nearest neighbor lattice sites that occur with acceptance probability one; that is, an ordinary random walk on a lattice. In this case, the Metropolis-Hastings moves perform a diffusion-like motion in phase space. 

We use the term ``diffusion'' for motion that requires $\sim N^2$ steps to travel a mean distance $N$ from its point of origin. What if this diffusive motion is too slow? We can imagine that sometimes it would be beneficial to have some ``inertia'' or ``momentum'' when performing auxiliary Markov chain hops in phase space, much like using a spoon to stir a cup of coffee helps to spread the sugar in the cup faster. Another familiar example is the smell of a cooked meal. If the odor molecules were only diffusing, they would reach across a dining table in a few hours instead of minutes. We can smell our meal in a timely way because of air currents. 

Markov chain Monte Carlo algorithms such as Metropolis-Hastings are especially slow close to phase transitions, where the dynamics suffers from critical slowing down due to large fluctuations of the observables. When sampling close to phase transitions, it is beneficial to introduce some inertia and bias.

The idea of ``lifting'' is to increase phase space to create a bias and explore the enlarged phase space more efficiently than we could explore the original space.\cite{CLP00, Diaconis:2000vi} Lifting alters the convergence time, and it is an open question if and when it will decrease the convergence time. The method we will introduce is potentially good for overcoming entropic barriers, but not for escaping deep energetic minima (see Sec.~\ref{sec:discussion}).

We can create lifting in an uncontrolled way by adding many cycles (by a cycle we mean here a closed walk -- a set of moves that starts and ends at the same point in phase space) because cycles in phase space do not change the steady state. The practical caveat is how many and what cycles to add to the already existing transitions. In Ref.~\onlinecite{turitsyn2011irreversible} we introduced a controlled way to create a nonreversible Markov Chain. Suppose that $\pi$ is a stationary distribution: $\pi = \pi P$, where $P$ is a stochastic matrix (\EQ{P_stochastic} holds). We define a larger space $\tilde\Omega = \Omega \times \{1,-1\}$ and denote a state in this space as $\{x_{\xi} \vert x\in \Omega, \xi \in\{1,-1\}\}$. Next we impose \emph{skew-detailed balance} 
\bal
\label{eq:SDB}
\tilde{\pi}(x_{\xi})\tilde{P}(x_{\xi},y_{\xi})=\tilde{\pi}(y_{-\xi})\tilde{P}(y_{-\xi},x_{-\xi})\,, 
\eal
for 
\bal
\label{eq:newstatdist}
\tilde{\pi} = \frac{1}{2}(\pi,\pi).
\eal
Recall that $\tilde{\pi}$ and $\pi$ are vectors. We enforce that the lifted transition matrix $\tilde{P}$ is stochastic,
\bal
\label{eq:Ptilde_stochastic}
\sum_{y_{\eta}\in\tilde{\Omega}}\tilde{P}(x_{\xi},y_{\eta})=1,~\forall x_{\xi}\in\tilde{\Omega},
\eal 
by adjusting the diagonal elements $\tilde{P}(x_{\xi},x_{\xi})=1 - \sum_{y_{\eta} \in \tilde \Omega:y_{\eta}\neq x_{\xi}}\tilde{P}(x_{\xi},y_{\eta})$.
The matrix $\tilde{P}$ has the following block structure: two diagonal blocks describe transitions inside $\Omega \times \{1\}$ and $\Omega \times \{-1\}$ spaces respectively; the off-diagonal blocks describe transitions between $x_\xi$ and $y_{-\xi}$ states. For simplicity, we assume that the off-diagonal blocks are diagonal matrices of the form 
\bal
\label{eq:OFFDIAG}
\tilde{P}(x_{\xi},y_{-\xi})=\delta_{xy}\tilde{P}(x_{\xi},y_{-\xi}).
\eal
A distribution $\tilde{\pi}$ satisfying the skew-detailed balance, \EQ{SDB}, is stationary with respect to $\tilde{P}$, that is, $\tilde{\pi} = \tilde{\pi}\tilde{P}$. We can prove this condition as follows: 
\bal 
\nonumber
& \sum_{y_{\eta}\in\tilde{\Omega}}\tilde{\pi}(y_{\eta})\tilde{P}(y_{\eta},x_{\xi}) 
=\sum_{y_{\xi}\in\tilde{\Omega}}\left[\tilde{\pi}(y_{\xi})\tilde{P}(y_{\xi},x_{\xi}) + \tilde{\pi}(y_{-\xi})\tilde{P}(y_{-\xi},x_{\xi}) \right].
\eal
We use skew-detailed balance, \EQ{SDB}, and \EQ{OFFDIAG} to obtain
\bal
\sum_{y_{\eta}\in\tilde{\Omega}}\tilde{\pi}(y_{\eta})\tilde{P}(y_{\eta},x_{\xi})=\tilde{\pi}(x_{-\xi})\left(\sum_{y_{\xi}\in\tilde{\Omega}}\tilde{P}(x_{-\xi},y_{-\xi}) + \tilde{P}(x_{-\xi},x_{\xi}) \right)
=\tilde{\pi}(x_{-\xi})\sum_{y_{\eta}\in\tilde{\Omega}}\tilde{P}(x_{-\xi},y_{\eta}) = \tilde{\pi}(x_{-\xi}),
\eal
where the last equality follows from \EQ{Ptilde_stochastic}. Finally, using \EQ{newstatdist} we obtain
\bal
& \sum_{y_{\eta}\in\tilde{\Omega}}\tilde{\pi}(y_{\eta})\tilde{P}(y_{\eta},x_{\xi}) = \tilde{\pi}(x_{\xi}),
\eal
which is the definition of stationarity and concludes our proof. 

We should also determine the off-diagonal elements. From stochasticity and \EQ{OFFDIAG} we have
\bal
\tilde{P}(x_\xi,x_{-\xi}) & = 1- \sum_{y_{\xi}\in\tilde{\Omega}}\tilde{P}(x_{\xi},y_{\xi}), \label{one} \\
\noalign{\noindent or}
\tilde{P}(x_{-\xi},x_{\xi}) & = 1- \sum_{y_{\xi}\in\tilde{\Omega}}\tilde{P}(x_{-\xi},y_{-\xi}). \label{two}
\eal
By subtracting Eqs.~\eqref{two} and \eqref{one} we obtain
\bal 
\tilde{P}(x_\xi,x_{-\xi})-\tilde{P}(x_{-\xi},x_\xi) 
=\sum_{y_\xi\in\tilde{\Omega}}\left(\tilde{P}(x_{-\xi},y_{-\xi})-\tilde{P}(x_{\xi},y_{\xi})\right).
\label{eq:inter_replica_transitions_diff}
\eal
From all possible solutions for $\tilde{P}(x_\xi,x_{-\xi})$ and $\tilde{P}(x_{-\xi},x_\xi)$, we want to choose the one for which these rates are minimal, because high rates impede the relaxation to equilibrium by fostering too many transitions between the two copies of the same state ($x_\xi$ and $x_{-\xi}$). The rates $\tilde{P}(x_\xi,x_{-\xi})$ and $\tilde{P}(x_{-\xi},x_\xi)$ are the smallest if one of them is zero, which leads to the following choice: 
\be
\label{eq:OFFDIAG_der}
\tilde{P}(x_{\xi},x_{-\xi}) \equiv {\max} \Big[0, \sum_{y_\xi \in\tilde{\Omega}}\left(\tilde{P}(x_{-\xi},y_{-\xi})-\tilde{P}(x_\xi,y_\xi)\right)\Big].
\ee
Note that there is still freedom in adjusting the $\tilde{P}(x_\xi,y_\xi)$ transition rates, even when the skew-detailed balance \EQ{SDB} is imposed -- this choice determines how much the detailed balance is violated.\cite{turitsyn2011irreversible,2013SakaiHukushima} 

\section{Applications of Lifting}
\label{sec:appl_lifting}
\subsection{Ring with a uniform stationary distribution}
\begin{figure}[t]
\includegraphics[width=0.65\columnwidth]{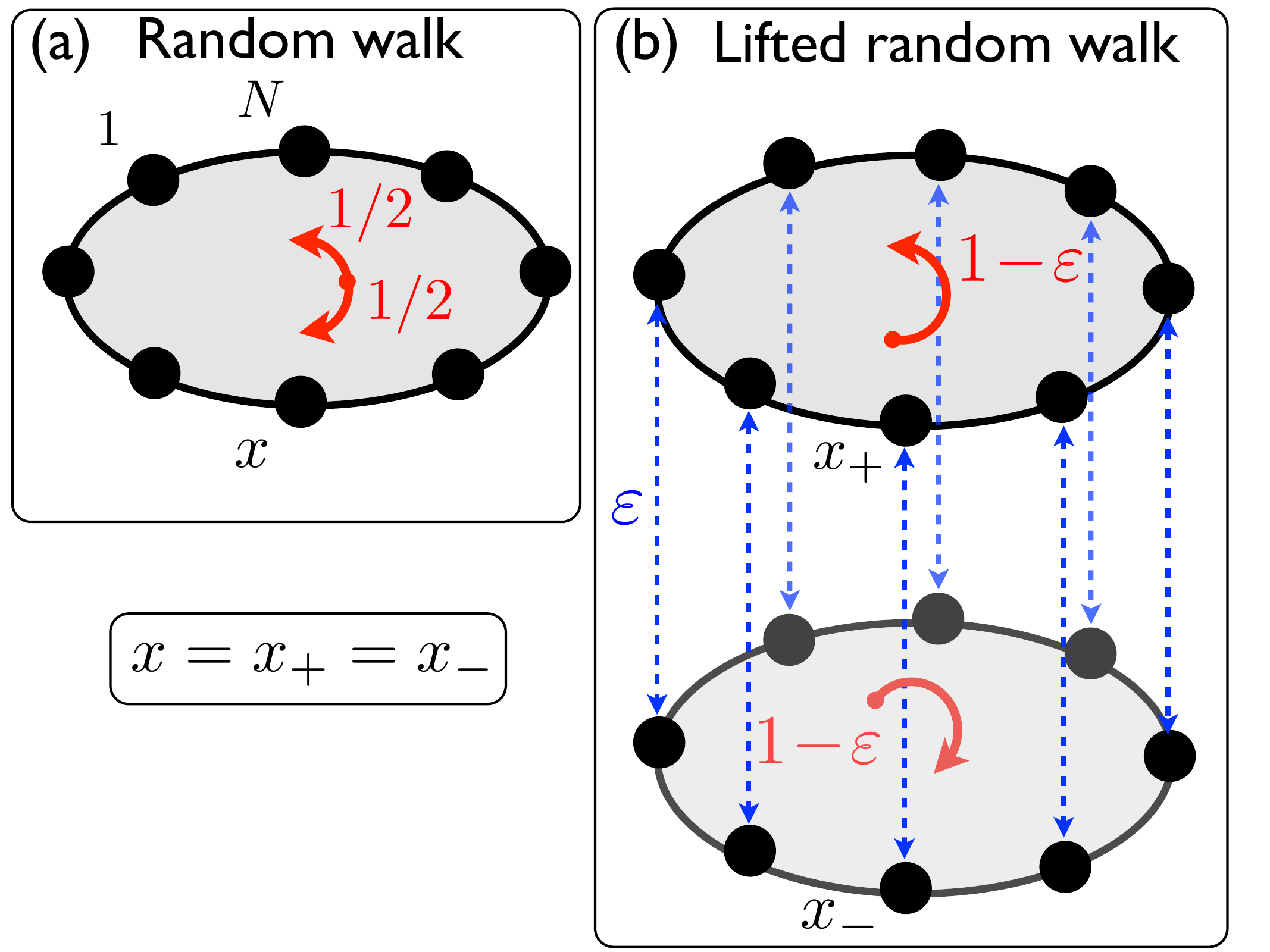}
\caption{\label{fig:N_cycle_v04.pdf} (a) Random walk on a ring with $N$ sites. The time to visit all sites is set by the diffusion time scale $t \propto N^2$. (b) Lifted random walk on a ring with $N$ states. We introduce two rings of $N$ sites. In the upper ring the moves have probability $1-\varepsilon$ counterclockwise and in the lower ring $1-\varepsilon$ clockwise, where $\varepsilon$ is small, for example, $\varepsilon \propto \mathcal{O}(N^{-1})$ for $N\to\infty$. The same state $x$ exists on both cycles as $x_+$ and $x_{-}$. These two copies of the state $x$ are connected by auxiliary transitions specified in \EQ{OFFDIAG_der} that ensure that global balance is maintained.}
\end{figure}

We first consider a Markov chain on a ring of $N$ states converging to a uniform distribution $\pi(x) = N^{-1}$ for all $x \in \{1, \ldots,N\}$. The idea is illustrated in \FIG{N_cycle_v04.pdf}. A random walker would cover every state along the ring in a time that scales as the diffusion time scale $t \sim N^2$ [see \FIG{N_cycle_v04.pdf}(a)]. Lifting can improve the convergence to the stationary distribution $\pi$. To apply lifting we create two rings of $N$ states: one on which transitions are made only in the counter clockwise direction and the other where transitions are made only in the clockwise direction. We set the bias $\varepsilon$ such that with probability $1-\varepsilon$ the walker continues to hop in the same direction; otherwise, the walker stays in the same state, but switches to the other replica of the system. This system converges to a steady state distribution $\pi$ after $\mathcal{O}(N)$ steps.\cite{Diaconis:2000vi,CLP00}

\subsection{Torus with uniform stationary distribution}
\begin{figure}
\includegraphics[width=\columnwidth]{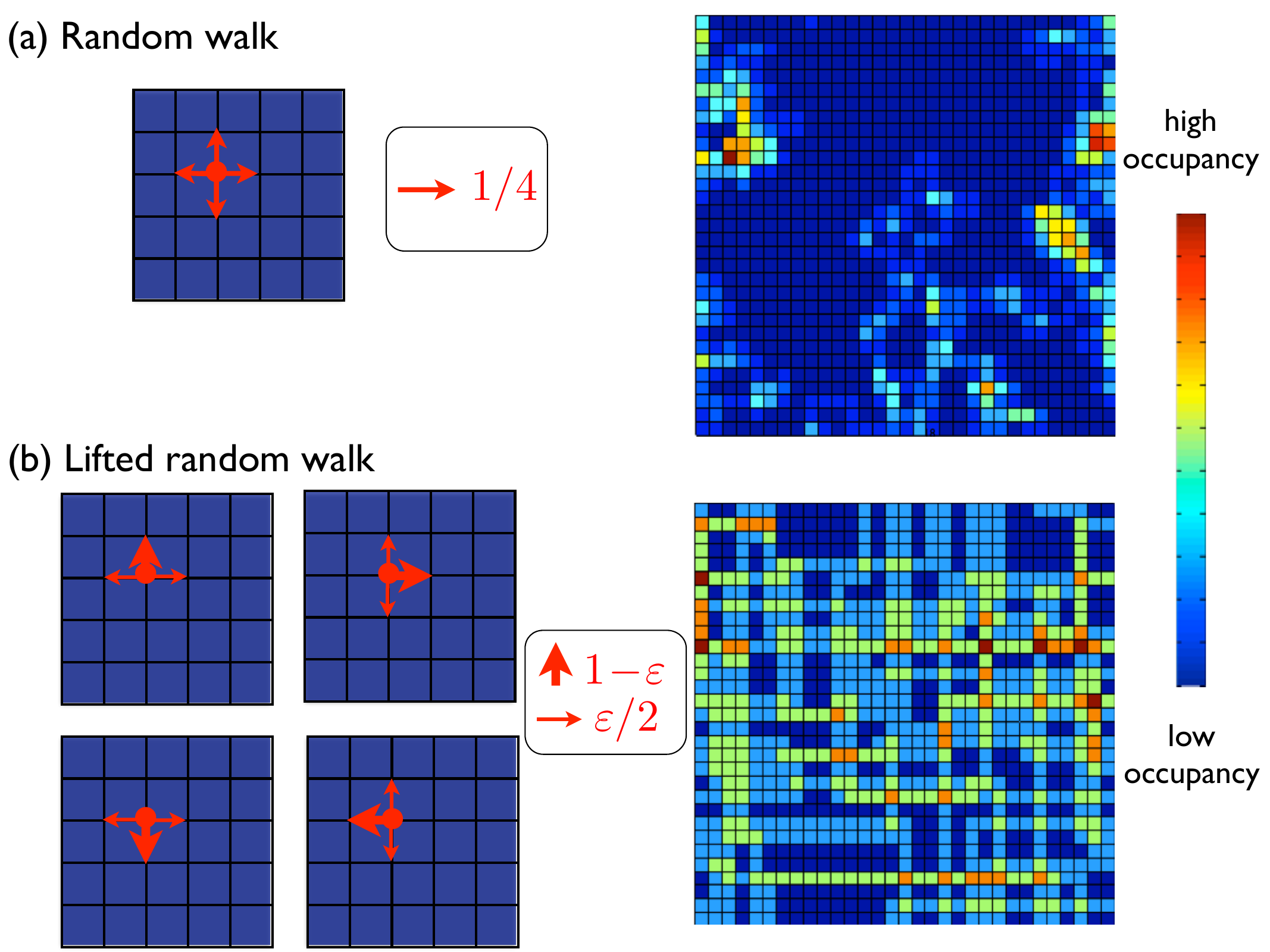}
\caption{\label{fig:torus_pic_v04.pdf} (a) (left) Random walk on a torus (square lattice with periodic boundary conditions). (right) Density of visited sites by a random walker on a torus of 1024 sites after 1024 steps. (b) (left) The torus is replaced by four tori with biased diffusion -- the north, south, east, and west tori. (right) Density of visited sites by a ``lifted'' random walker projected back on the original torus of 1024 sites after 1024 steps. The bias is $\varepsilon = 0.1$. Notice the long vertical and horizontal strides that the lifted random walker makes while exploring the tori.}
\end{figure}

\begin{figure}
\includegraphics[width=\columnwidth]{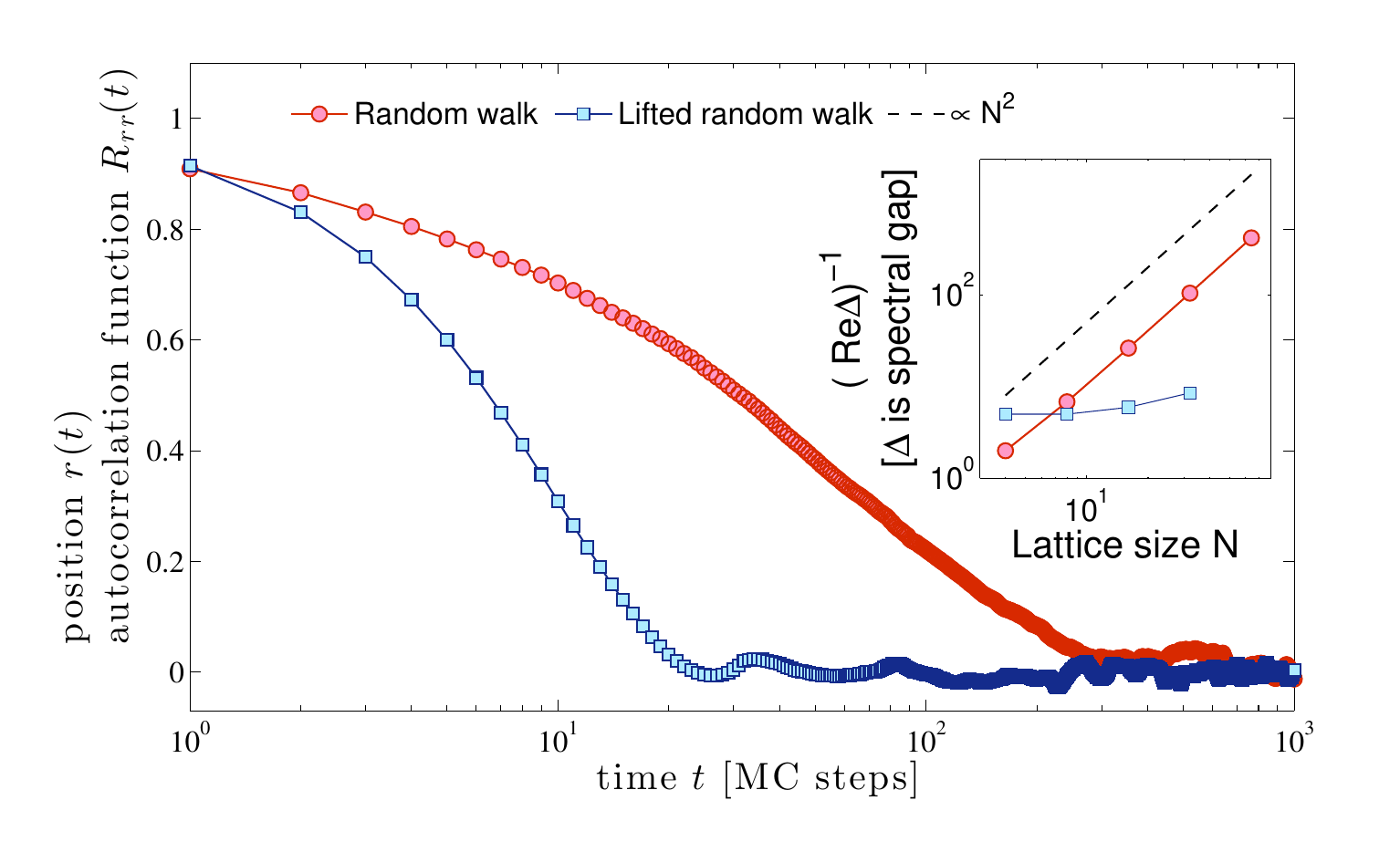}
\vspace{-0.3in}
\caption{\label{fig:autocorrelationfunction_spectral_gap_torus_v06.pdf}Decay of the position $r(t)$ autocorrelation function $R_{rr}(t)$ [see \EQ{autocorr_norm}], indicating the approach to a steady state. (inset) Dependence of the real part of the inverse spectral gap on the lattice size $N$ for a torus with $N^2$ sites. Note that the autocorrelation function of the lifted random walk decays more rapidly than the autocorrelation function of the (unbiased) random walk. Likewise, the inverse of the real part of the spectral gap increases more slowly for the lifted random walk. Both of these characteristics indicate faster relaxation to equilibrium.}
\end{figure}

A reversible Markov chain specified with an arbitrary initial distribution $\rho(t = 0,x)$, and a transition matrix with equally likely transitions to neighboring sites on a square lattice with periodic boundary conditions and $N^2$ sites, converges to a uniform stationary distribution $\pi(x) = N^{-2}$ (for $x \in \{1,\ldots,N^2\}$) after $\mathcal{O}(N^{2})$ steps (see for example, Ref.~\onlinecite{CLP00} or the inset in \FIG{autocorrelationfunction_spectral_gap_torus_v06.pdf}). Such a Markov chain can be visualized as a random walk with transition probability $\varepsilon = 1/4$ to any of the four nearest neighbor sites on a square lattice spanning a torus [see \FIG{torus_pic_v04.pdf}(a)]. One way to define a nonreversible Markov Chain on this torus is to give the random walker some ``inertia'' and define the walk as follows: if the walker enters a site from a particular direction it will exit continuing in the same direction with probability ($1-\varepsilon$), or it will turn left with probability $\varepsilon/2$, or turn right with probability $\varepsilon/2$, but it will never return to the site that it just came from [see \FIG{torus_pic_v04.pdf}(b)]. That is, if the walker started its walk in a given direction, say toward north (and we defined north, south, east and west on our torus), it is likely to continue going north until it walks about $2 N /\varepsilon$ steps. The site where it will eventually turn and change direction (to walk east or west) is uniformly distributed over the south-north axis, because the walker has made roughly $2/\varepsilon$ circles around the torus walking always north. Next, having chosen to walk either toward east or west, it continues along the same direction for another approximately $2 N /\varepsilon$ steps. When it will turn again, its east-west (or $x-$coordinate) will also be uniformly distributed. Hence by the second turning point the position of the walker will be uniformly distributed over all sites of the lattice. Hence, after about $4 N / \varepsilon$ steps the walker is equally likely to be on any site on the torus in a random realization of this process. Note that to reach this second turning point, the walker needed only $\mathcal{O}(N)$ steps. This idea is described in Ref.~\onlinecite{CLP00}, and shown in Figs.~\ref{fig:torus_pic_v04.pdf} and \ref{fig:autocorrelationfunction_spectral_gap_torus_v06.pdf}. Instead of diffusing on a single torus, the walker, depending on its direction, walks on one of the four tori represented in \FIG{torus_pic_v04.pdf}. Every time when the walker turns, the torus on which it walks changes, so that the most probable step is always along the same direction as the previous step of the walker. The decay of the pair correlation function and the scaling of the real part of the inverse gap with the system size is shown in \FIG{autocorrelationfunction_spectral_gap_torus_v06.pdf}. Both of these measures of convergence imply that the lifted random walk converges faster than the unbiased random walk. The autocorrelation function of the lifted random walk decays faster (decorrelates more rapidly). Likewise, the inverse of the real part of the spectral gap increases more slowly for the lifted random walk indicating a faster relaxation to equilibrium. 

\subsection{The Ising model on a complete graph}
Consider $N$ Ising spins on a complete graph (every pair of distinct vertices is connected by a unique edge). This system is also known as the fully connected Ising model because every spin interacts with every other spin. The system exhibits a continuous phase transition, symmetry breaking, and emergence of spontaneous magnetization at a nonzero positive temperature in the limit as the number of spins $N\to \infty$. Let each vertex carry a spin $\sigma_i \in\{1,-1\}$. The energy of a spin configuration $\bm \sigma = \{\sigma_1,\ldots,\sigma_N\}$ is 
\bal
E(\bm \sigma) = - \frac{J}{N}\sum_{i<j}\sigma_i\sigma_j, 
\eal 
and the sum runs over all pairs of spins. The ground states correspond to all spins $\sigma_i$ pointing up ($\sigma_i = 1$) or all spins pointing down ($\sigma_i = -1$), that is, the ground state is degenerate with entropy $S = \ln 2$. The ground state energy is $E_g = - (J/2)(N-1)$. Note that $E_g \sim \mathcal{O}(N)$, while $S_g \sim \mathcal{O}(1)$. At $T = \infty$ all states are equally probable giving rise to $\langle E \rangle = 0$, and $S =N \ln 2$, because the number of configurations is $Z = 2^N$. Note that at $T=\infty$ the entropy scales with the system size $S\propto \mathcal{O}(N)$. At high temperatures the spins are disordered and the average magnetization is zero, while at low temperatures the spins tend to align with a magnetization $M = \sum_{i=1}^N \sigma_i$, which is nonzero (although the average magnetization remains zero). An infinitesimal perturbation will determine which ground state is selected. At some $T> 0$, where the energy and entropy are of the same order of magnitude, there is a phase transition.

\begin{figure}[t]
\includegraphics[width=3.42in]{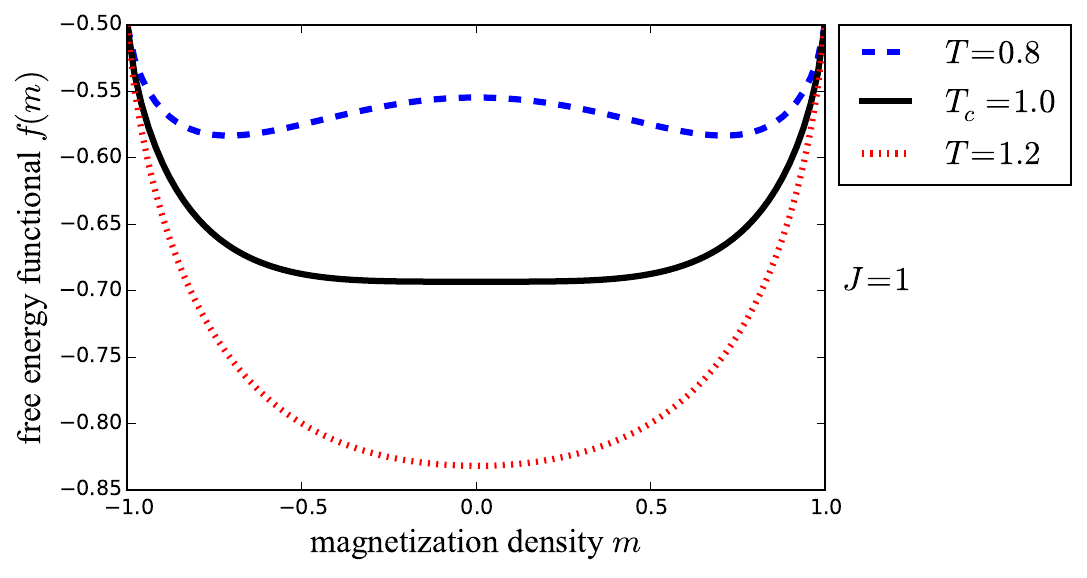}
\caption{\label{fig:free_energy_complete_graph_ising.pdf} The free energy functional for $J = 1$ and three temperatures. $T_c = J$ is the critical temperature. Above $T_c$ the probability distribution of the magnetization is centered around 0, and below $T_c$ there are are two nonzero minima. The degeneracy between the two is lifted by a small perturbation and remains lifted even after the perturbation has vanished. At $T_c$ the curvature of the saddle point $m =0$ vanishes, which signifies a continuous phase transition.}
\end{figure}

A key observation in this model is that the energy depends solely on the magnetization: $E =-(J/2N)(M^2-1)$. Instead of summing over configurations, the partition function can be written as a sum of $N+1$ terms, the number of different values of magnetization. Each magnetization occurs with multiplicity $D(M) =\binom{N}{N_+}$, where $N_+ = (N+M)/2$ is the number of positive spins. Therefore the partition function is 
\bal
Z(T) = \sum_{\bm \sigma}e^{-\beta E(\bm \sigma)} = \sum_{m=-1, \ldots,1}D(m)e^{\beta E(m)}, 
\eal
where $m = M/N$. The entropy density $s = S/N$ at fixed magnetization, in the limit of large $N$, is 
\bal
\frac{S}{N} = - \frac{1+m}{2}\ln \Big(\frac{1+m}{2}\Big)-\frac{1-m}{2}\ln 
\Big(\frac{1-m}{2}\Big).
\eal
Because $s = N \ln D(m)$, we have, in the limit of large $N$,
\bal
Z &= \sum_{m=-1, \ldots,1}e^{-\beta N f(m)}, \\
\label{eq:free_energy_functional}
f(m) &= - \frac{J}{2}m^2 - Ts(m), 
\eal
where $f(m)$ is the free energy functional. As seen in \FIG{free_energy_complete_graph_ising.pdf}, there is a critical temperature $T_c = J$, below which there are two free energy minima at $m\propto\mathcal{O}(1)$, and above which there is only one free energy minimum at zero magnetization. The degeneracy between the two free energy functional minimums is lifted with a small perturbation (such as an external magnetic field) and remains lifted even after the perturbation has vanished. As $N\to\infty$ there is a phase transition, but at finite $N$ there are magnetization fluctuations proportional to $\mathcal{O}(N^{\delta-1})$, where $\delta$ is given in the following. If we expand the free energy functional \EQ{free_energy_functional} for $T$ close to $T_c$, we have
\bal
f(m) \approx \frac{\tau}{2}m^2 +\frac{m^4}{12} -T\ln 2 + \mathcal{O}(m^6,\tau m^4),
\eal
with $\tau = 1 - T/T_c$. For $\tau >0$ the fluctuations of $M$ are proportional to $N^{-1/2}$, which gives $\delta = 1/2$. At the critical temperature the quadratic term and the average magnetization vanish, but the fluctuations are of order $N^{-1/4}$, and thus $M$ has a distribution of width $N^{3/4}$ (see Ref.~\onlinecite{PhysRevE.90.042111}). Reference~\onlinecite{2010GouldTobochnik} gives an excellent introduction to spin systems and their simulation.

\begin{figure}[t]
\includegraphics[width=3.42in]{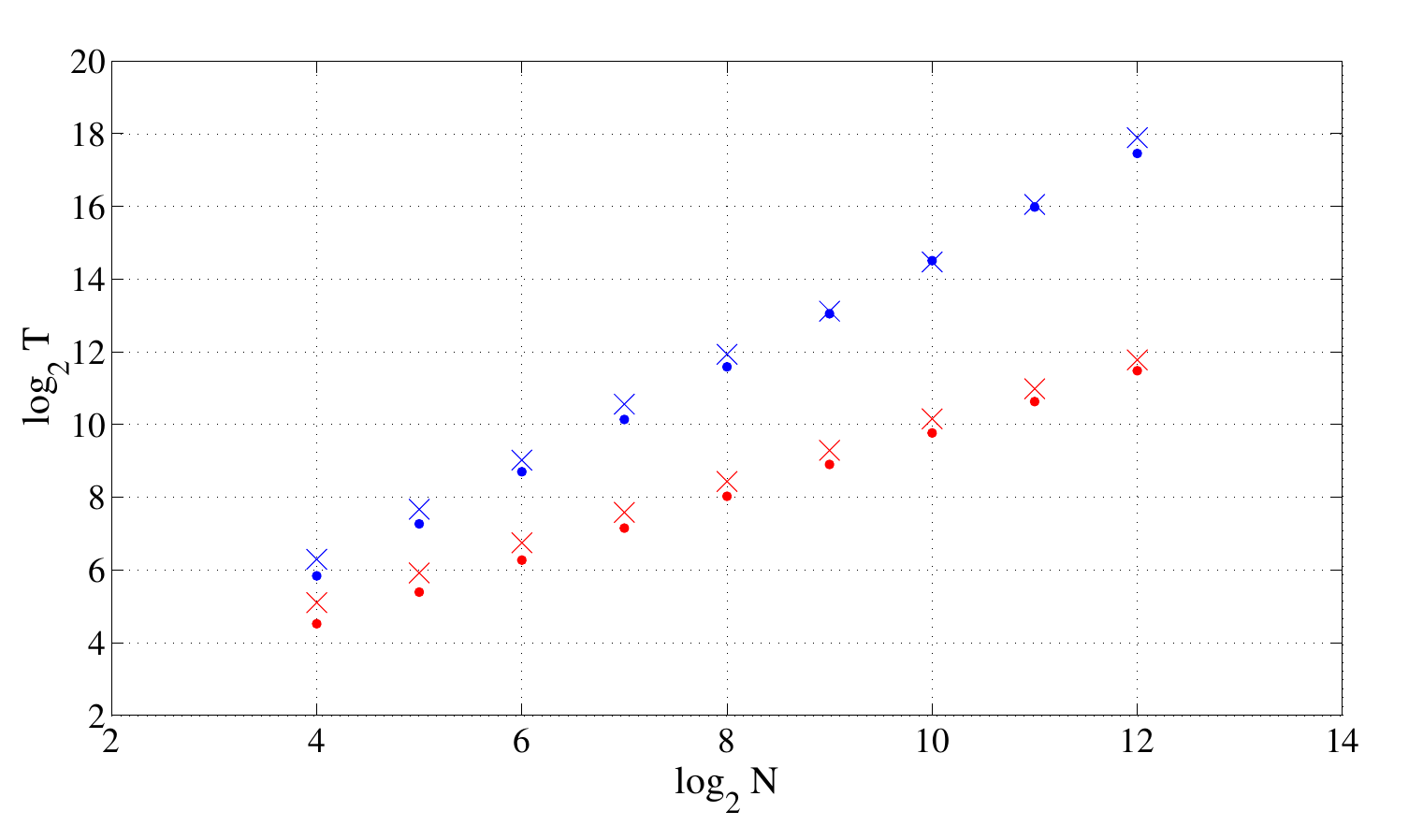}
\caption{\label{fig:mixingtime.pdf} (Color online) The correlation time of the magnetization autocorrelation function of the Ising model on the complete graph. The blue data at the top represent the reversible Metropolis-Hastings algorithm and the red data points at the bottom are from the nonreversible Metropolis-Hastings algorithm. The dots represent $T = 1/[{\rm Re}(\Delta)]$, where $\Delta$ is the spectral gap, obtained by exact diagonalization of the respective transition matrices. The crosses represent correlation times reconstructed by fitting the large time behavior by $\exp(-t /\tau_{\rm rev})$, and exponential-oscillatory function, $\exp(-t/\tau_{\rm nonrev})\cos(\omega t -\phi)$, for the reversible and nonreversible cases, respectively. The best slope fits are $\tau_{\rm rev}(N)\propto N^{1.43}$ and $\tau_{\rm nonrev}(N)\propto N^{0.85}$. Reproduced from Ref.~\onlinecite{turitsyn2011irreversible}.}
\end{figure}

The time it takes for a reversible Markov chain Monte Carlo algorithm to decorrelate $R_{mm}(t)$ is proportional to the variance of $m$, that is, $\propto N^{3/2}$ close to the critical point. In contrast, the proposed lifting algorithm converges as $N^{3/4}$. We introduced the algorithm and confirmed $N^{3/4}$ numerically in Ref.~\onlinecite{turitsyn2011irreversible} by measuring the inverse spectral gap $\Delta^{-1}$ and the decay of the autocorrelation function of the magnetization $R_{mm}(t)$. These numerical results are reproduced in \FIG{mixingtime.pdf}. In addition, the $N^{3/2}$ and $N^{3/4}$ scalings were recently rigorously proven in Ref.~\onlinecite{2015arXiv150900302B}. 

The lifting algorithm from Ref.~\onlinecite{turitsyn2011irreversible} goes as follows: We create two copies of the system. The two copies always have the same spin configuration; it is just the transition rates in the two copies that are different. In the $(+)$ copy we prefer to flip $\uparrow$ spins, which will decrease the magnetization and in the $(-)$ copy we prefer to flip $\downarrow$ spins, which will increase the magnetization. Let us assume that the system is initially in state $x_+$ (belongs to the $(+)$ copy). Next we randomly select an $\uparrow$ spin. We try a Metropolis-Hastings move to flip the chosen spin (a flip that would change the state from $x_+$ to $y_+$). The move is accepted with probability $a(x_+,y_+)= {\min}\left[1, \frac{\pi(y_+)}{\pi(x_+)}\right]$. If the move is rejected, then with probability $q$ (the explicit expression of $q$ is given in \FIG{lifting}) we change the copy from $(+)$ to $(-)$. If both moves are rejected, the system stays at the same state and in the same copy of the system and we choose another $\uparrow$ spin and repeat the outlined steps. See the pseudocode in \FIG{lifting}. This algorithm results in an effective magnetic field that depends on the state of the system and allows the system to linger longer at states of very low and very high magnetization.\cite{turitsyn2011irreversible} Similar observations were made for the one-dimensional Ising model in Ref.~\onlinecite{2013SakaiHukushima}. Ultimately the lifted Markov chain converges faster to equilibrium than the corresponding reversible Markov chain.\cite{turitsyn2011irreversible}

%\begin{algorithm}[t]
\begin{figure}[t]
\noindent\makebox[\linewidth]{\rule{\columnwidth}{0.4pt}}
\begin{algorithmic}
	\STATE\COMMENT {initially: system at state $x_+$}
	\STATE $s_i\gets{\text{randomly selected} \uparrow \text{spin from } x_+}$
	\STATE $a(x_+,y_+) = {\min}\left[1, \frac{\pi(y_+)}{\pi(x_+)}\right]$\COMMENT {Metropolis-Hasting algorithm (flipping $s_i$ in $x_+$ gives $y_+$)}
	\STATE $p \gets{\text{random number in }(0,1)}$ 
	\IF {$p \leq a$}
 	\STATE spin is flipped and the new state is $y_+$ \COMMENT{Metropolis-Hasting move}
	\ELSE \COMMENT{attempt to change from $x_+$ to $x_-$}
		\STATE $q\gets{\text{random number in }(0,1)}$ 
		\IF {$q \leq \frac{\tilde{P}(x_+,x_-)}{(1 - \sum_{z_+;z_+\neq x_+} \tilde{P}(x_+,z_+))}$}
 	\STATE the new state is $x_-$ 
 	\ENDIF
	\ENDIF
\end{algorithmic}
\noindent\makebox[\linewidth]{\rule{\columnwidth}{0.4pt}}
\caption{Implementation of lifting using a Metropolis-Hastings Markov chain Monte Carlo algorithm.}
\label{fig:lifting}
\end{figure}
%\end{algorithm}

\subsection{One-dimensional Ising model}
Consider a one-dimensional Ising model of $N$ spins with periodic boundary conditions and nearest neighbor interaction $J$. A state of the system is a spin configuration $\bm \sigma = \{\sigma_1, \ldots,\sigma_N \}$, and the energy of this configuration is 
\bal
E(\bm \sigma) = - J \sum^{N-1}_{i=1}\sigma_i\sigma_{i+1}. 
\eal 
The system is connected to a thermal reservoir (heat bath), which is kept at temperature $\beta^{-1}$. The equilibrium probability for the system to be in a state $\bm \sigma$ is the Gibbs distribution
\bal
\pi(\bm \sigma) \propto e^{-\beta E(\bm \sigma)}.
\eal
Our goal is to sample equilibrium properties of this system, but once again direct sampling of $\pi (\bm \sigma)$ is unfeasible, because 
that system can be in exponentially many states ($2^N$). One way to relax the system to equilibrium is to use a Monte Carlo algorithm with heat bath acceptance 
probabilities (see Sec.~\ref{sec:mh}).

It is useful to define a spin-flip operator $F_j$. This operator acts on the spin configuration $\bm \sigma$ by flipping only the $j$th spin:
\bal
\label{eq:spin_flip_operator}
F_j\bm \sigma = F_j \{\sigma_1, \ldots, \sigma_j, \ldots, \sigma_N\} = \{\sigma_1, \ldots, - \sigma_j, \ldots, \sigma_N \}. 
\eal
The heat bath acceptance probability for flipping the $j$th spin is given by \EQ{heat_bath_acceptance_probability}: 
\bal
\label{eq:1d_Ising_heat_bath_acceptance_probabilities}
a(\bm \sigma, F_j \bm \sigma) = \frac{\pi (F_j\bm \sigma)Q (F_j \bm \sigma, \bm \sigma)}{\pi(F_j\bm \sigma) Q(F_j \bm \sigma,\bm \sigma) + \pi(\bm \sigma)Q(\bm \sigma, F_j \bm \sigma)}.
\eal
The initial transition matrix, $Q$, is inconsequential (as we argued in Sec.~\ref{sec:mh}), and in order to make the algorithm simpler, we choose it to be symmetric: 
\bal
Q(\bm \sigma, F_j \bm \sigma) = Q (F_j \bm \sigma, \bm \sigma).
\eal 
In this case \EQ{1d_Ising_heat_bath_acceptance_probabilities} reduces to
\bal
\label{eq:1d_Ising_heat_bath_acceptance_probabilities_v01}
a(\bm \sigma, F_j \bm \sigma) = \frac{\pi (F_j\bm \sigma)}{\pi(F_j\bm \sigma) + \pi(\bm \sigma)}.
\eal
For the one-dimensional Ising model $a(\bm \sigma, F_j \bm \sigma)$ reduces to 
\bal
\label{eq:glauber_1d_DBC}
a(\bm \sigma, F_j \bm \sigma) = \frac{\alpha}{2} \left[ 1 - \frac{\gamma}{2} \sigma_j (\sigma_{j-1}+\sigma_{j+1})\right],
\eal
where $\gamma \equiv \tanh(2 J \beta)$ and $\alpha$ sets the unit of time. This kind of stochastic dynamics was first introduced by Glauber.\cite{Glauber} Glauber dynamics and the heat bath acceptance probabilities happen to be identical for the Ising model (in other physical systems the two are distinct). 

By direct substitution of $a(\bm \sigma, F_j \bm \sigma)$ and $Q$, we can verify that the detailed balance condition
\bal
\pi (\bm \sigma) P(\bm \sigma,F_j \bm \sigma) = \pi (F_j \bm \sigma) P(F_j \bm \sigma, \bm \sigma) 
\eal
holds. 

Sakai and Hukushima implemented lifting for this problem.\cite{2013SakaiHukushima} Their lifted Markov chain Monte Carlo for the one-dimensional Ising model has the following heat bath acceptance probabilities
\bal
\label{eq:glauber_1d_SDB}
\tilde{a}(\bm \sigma _\xi, F_j \bm \sigma _\xi ) = \frac{\alpha}{2}\left[1 - \frac{\gamma}{2}\sigma_j \gamma \left(\sigma _{j-1} + \sigma _{j+1}\right)\right]\left(1 - \delta \xi \sigma _j\right),
\eal
where $\delta \in [-1,1]$ quantifies how much detailed balance is violated (for $\delta = 0$ there is no violation of detailed balance: $a = \tilde{a}$).

The new transition matrix:
\bal
\label{eq:Ptilde_1d_SDB}
\tilde{P}(\bm \sigma _\xi, F_j \bm \sigma _\xi) = \tilde{a} (\bm \sigma _\xi, F_j \bm \sigma _\xi) \tilde{Q}(\bm \sigma _\xi, F_j \bm \sigma _\xi),
\eal
obeys skew detailed balance [see \EQ{SDB}]: 
\bal
\label{eq:SDB_1d_Ising}
\tilde{\pi}(\bm _{\bm \sigma _\xi})\tilde{P}(\bm \sigma _\xi, F_j \bm \sigma _\xi)=
\tilde{\pi}(\bm _{\bm \sigma _{-\xi}})\tilde{P}(\bm \sigma _{-\xi}, F_j \bm \sigma _{-\xi})\,.
\eal
Here we assume that transition matrix $\tilde{Q}$ is symmetric. Next, \EQ{inter_replica_transitions_diff} specifies the difference between 
inter replica transitions (transitions between $\bm \sigma _\xi$ to $\bm \sigma _{-\xi}$)
\bal
\tilde{P}(\bm \sigma _\xi, \bm \sigma _{-\xi}) -\tilde{P}(\bm \sigma _{-\xi}, \bm \sigma _{\xi}) = \sum _{\bm \sigma' \in \Omega} \left(
\tilde{P}(\bm \sigma _{-\xi}, \bm \sigma' _{-\xi}) -\tilde{P}(\bm \sigma _{\xi}, \bm \sigma' _{\xi}) \right). \label{above}
\eal
Sakai and Hukushima examined three different solutions for Eq.~\eqref{above}, one of which was \EQ{OFFDIAG_der}, and concluded that all 
three nonreversible Markov chain Monte Carlo algorithms converge faster that the corresponding reversible Markov chain Monte Carlo.\cite{2013SakaiHukushima} 

They also made a very insightful remark that the chosen rate transition $\tilde P$ obeys detailed balance \EQ{DBC} for a one-dimensional Ising model in a particular magnetic field $H = \arctanh(- \delta \xi)/\beta \xi$, with energy: 
\bal
E(\bm \sigma) = - J \sum _{j = 1}^{N-1} \sigma_j \sigma_{j+1} + \xi H \sum _{j = 1} ^N \sigma _j.
\eal
In other words, the lifted Markov chain Monte Carlo for the one-dimensional Ising model has transition rates equal to those of a reversible Markov chain Monte Carlo algorithm for a one-dimensional Ising model in a magnetic field that depends on the state of the system. The skew-detailed balance \EQ{SDB_1d_Ising} condition ensures that the nonreversible Markov Chain converges to the equilibrium distribution without the magnetic field.\cite{2013SakaiHukushima} The net effect of this virtual magnetic field seems to make the lifted Markov chain Monte Carlo algorithm faster than it's reversible counterpart.

\subsection{Two-dimensional Ising model -- caveats}
An application of lifting, similar to the previous two examples, by controlling the magnetization, does not yield a significant speedup for the two-dimensional Ising model at the critical point.\cite{Ferna2011} Likewise lifting, by creating two replicas where instead of the magnetization the total energy is controlled (in one replica the system can only increase its energy and the other it can only decrease its energy), does not lead to a significant speedup either.\cite{Ferna2011} In all of these cases adding nonreversible moves seems to affect only the numerical pre-factor of the convergence time, but not the scaling with the system size.\cite{Ferna2011, 2013SakaiHukushima} 

Further investigation is needed on how to make lifting adaptive to the energy barriers and low entropy paths of the configuration space. Although we have successfully created a algorithm that obeys global balance and converges to the proper equilibrium distribution, we have not yet been able to find the lifting algorithm that leads to significantly faster convergence for the two-dimensional Ising model.

The lifting that provides the fastest convergence has to utilize the physics of the system. For example, in the mean-field Ising model the slowest observable to converge is the magnetization and the equilibrium distribution can be written as a function of only the magnetization. Thus it is natural to choose a lifting that introduces a bias in the way the magnetization is sampled. 

\section{Discussion and comparison to other methods}
\label{sec:discussion}
Besides the mentioned implementations of lifting,\cite{Ferna2011, 2013SakaiHukushima} there are several other similar ideas.\cite{2013Ichiki,PhysRevLett.105.120603,2009BKW,2012BK,Schram201588} Suwa and Todo have proposed a Markov chain Monte Carlo method that violates detailed balance and reduces the convergence time of the model compared to the corresponding reversible Markov chain Monte Carlo methods.\cite{PhysRevLett.105.120603} Their algorithm minimizes the average rejection rate (probability of rejecting a proposed move) and requires summing over all states. For the two-dimensional Ising model of 16 spins, the Suwa-Todo algorithm has an integrated autocorrelation time that is 6.4 times shorter than the Metropolis-Hastings algorithm.\cite{PhysRevLett.105.120603} It would be very interesting to see a study of convergence rates for the Suwa-Todo algorithm as a function of the number of spins. A new nonreversible algorithm was developed for hard-sphere systems,\cite{2009BKW,2012BK} with substantial acceleration compared to related variants obeying detailed balance. For a more detailed comparison of the algorithms that violate detailed balance we refer readers to Ref.~\onlinecite{turitsyn2011irreversible}.

More rigorous results pertinent to our work can be found in Refs.~\onlinecite{Diaconis:2000vi,CLP00,Hayes,2014arXiv1401.8087B,09GB,2013lelievre,2014arXiv1404.0105R}. Much less is known about nonreversible Markov chains compared to the vast knowledge of reversible Markov chains (see, for example, Ref.~\onlinecite{2009Levinbook}). For example, the Peskun theorem holds for reversible Markov chains. This theorem states that the asymptotic variance of any observable is reduced by increasing the acceptance probability of the Markov chain ($a(x,y)$ in our notation for the Metropolis-Hastings Markov chain). Reference~\onlinecite{CLP00} shows that lifting can at most introduce a square root improvement of the convergence time. Such an acceleration is still quite impressive for long convergence times.

We have discussed how to controllably transform a reversible Markov chain Monte Carlo algorithm into a nonreversible Markov chain Monte Carlo algorithm for several models. The main idea is to enlarge the phase space to facilitate easier ``escape'' of entropic bottlenecks. This method is not designed for efficient sampling of rugged energy landscapes, where the convergence time to equilibrium is determined by rare events that escape deep energy wells. Lifting is potentially useful where there are entropic barriers -- such as a vast energetically almost flat configuration space or a maze with paths of low entropy. Lifting does not require any particular symmetry of the configuration space (for example, it does not rely on the $\mathbb{Z}_2$ symmetry of the Ising model). Our simple examples show that lifting can lead to a dramatic reduction of the convergence time. Methods using non-equilibrium mixing (methods that violate detailed balance) might prove useful in studies of phase transitions, soft matter dynamics, protein structures, and granular media. An interesting direction for future research is to explore if the lack of reversibility can improve the convergence properties of well known reversible algorithms.

\section{Suggested problems}
\label{sec:sugg_problems}

\noindent Problem 1. {\it Two walkers on a torus}. Assume that we place two walkers on a square lattice with periodic boundary conditions. The initial distance between the two walkers is $r_0$ and there are $N^2$ sites on this torus. Arbitrarily define north, south, east and west on this torus. What is the average distance between the two walkers after each of the two walkers has taken $t$ steps?
\begin{itemize}
\item[(a)]Assume that both of the walkers perform an unbiased random walk with the transition probabilities
\bal
P(x,y) = 
\begin{cases}
1/4 & \text{for nearest neighboring lattice sites $x$ and $y$}\\
0& \text{otherwise}
\end{cases}
\eal 

\item[(b)]Assume that one of the walkers performs an unbiased random walk as before, while the other performs a walk with some inertia ($\varepsilon \neq 0$). Its transition matrix $Q(x,y)$ is
\bal
Q(x,y) = \begin{cases}
1 - \varepsilon, & \text{if $x \to y$ points toward the north; $x$ and $y$ are nearest neighboring lattice sites}\\
\varepsilon/2, & \text{if $x \to y$ points toward the east; $x$ and $y$ are nearest neighboring lattice sites}\\
\varepsilon/2, & \text{if $x \to y$ points toward the west; $x$ and $y$ are nearest neighboring lattice sites} \\
0, &\text{otherwise}.
\end{cases}
\eal

\end{itemize}
Express the average distance between the two walkers after time $t$ as a function of $\varepsilon$.

\medskip \noindent Problem 2. {\it Heat bath acceptance probability for the one-dimensional Ising model}. Start from \EQ{1d_Ising_heat_bath_acceptance_probabilities_v01} and derive \EQ{glauber_1d_DBC} for the one-dimensional Ising model. 

\medskip \noindent Problem 3. {\it The lifted transition matrix for the one-dimensional Ising model}.
Show that $\tilde P (\bm \sigma _\xi, F_j \bm \sigma_\xi)$, defined in \EQ{Ptilde_1d_SDB}, satisfies the detailed balance condition for a one-dimensional Ising model in the magnetic field $H=\arctanh(- \delta \xi)/\beta \xi$. 

\medskip \noindent Problem 4. {\it One-dimensional Ising model}.
Write a program to implement the nonreversible Metropolis-Hastings algorithm by following the pseudocode in \FIG{lifting} for the one-dimensional Ising model. Compare your results with those of the conventional Metropolis-Hastings algorithm (using reversible Markov chains). 

\begin{acknowledgments}
Part of this work was completed at the Aspen Center for Physics. I acknowledge the Aspen Center for Physics and NSF grant $\#1066293$ for support. I thank J.\ Machta, K.\ S.\ Turitsyn, M.\ Chertkov, C.\ Moore, K.~Hukushima, W.\ Krauth, C.\ Godr\' eche, T.\ Hayes and J.\ Bierkens for illuminating discussions. 
\end{acknowledgments}

\end{document}